\documentclass[twocolumn]{aastex63}

\usepackage{mathrsfs}
\usepackage[title]{appendix}

\usepackage{lineno}

\usepackage{amsmath}
\usepackage{enumitem}
\usepackage{hyperref}
\hypersetup{colorlinks,linkcolor={blue},citecolor={blue},urlcolor={blue}} 

\def\mbh{\textit{$\mathcal{M}_{\rm BH}$}}
\def\m{\textit{$\mathcal{M}_{\star}$}}

\def\muv{\textit{$M_{\rm 1450}$}}

\def\mgas{\textit{$\mathcal{M}_{\rm gas}$}}
\def\mdy{\textit{$\mathcal{M}_{\rm dyn}$}}

\def\mbulge{\textit{$\mathcal{M}_{\rm bulge}$}}
\def\logm{\textit{${\rm log}\,\m$}}
\def\logmbh{\textit{${\rm log}\,\mbh$}}

\def\deltalogmbh{\textit{$\Delta\,\logmbh$}}

\def\rel{\textit{$\mbh-\m$}}
\def\bulgerel{\textit{$\mbh-\mbulge$}}

\def\hb{\textit{${\rm H}\beta$}}

\def\cii{{\rm [C~\scriptsize{II}}]}

\def\mgii{{\rm Mg~\scriptsize{II}}}
\def\civ{{\rm C~\scriptsize{IV}}}

\def\msun{\textit{$M_\odot$}}
\def\ergs{\textit{$\rm erg\ s^{-1}$}}

\def\gm{\textit{$\gamma$}}

\def\lbol{\textit{$L_{\rm bol}$}}
\def\sigmamu{\textit{$\sigma_\mu$}}

\def\nuw{\textit{${\rm FWHM}_{\mgii}$}}

\def\edd{\textit{$\lambda_{\rm Edd}$}}

\def\logedd{\textit{${\rm log}\,\lambda_{\rm Edd}$}}

\graphicspath{{./}{figures/}}

\begin{document}

\title{On the Connection between Supermassive Black Hole  and Galaxy Growth in the Reionization Epoch}

\correspondingauthor{Junyao Li}
\email{lijunyao@mail.ustc.edu.cn}

\author{Junyao Li}
\affiliation{CAS Key Laboratory for Research in Galaxies and Cosmology, Department of Astronomy, University of Science and Technology of China, Hefei 230026, China}
\affiliation{Kavli Institute for the Physics and Mathematics of the Universe (WPI), The University of Tokyo, Kashiwa, Chiba 277-8583, Japan}
\affiliation{School of Astronomy and Space Science, University of Science and Technology of China, Hefei 230026, China}

\author{John D. Silverman}
\affiliation{Kavli Institute for the Physics and Mathematics of the Universe (WPI), The University of Tokyo, Kashiwa, Chiba 277-8583, Japan}
\affiliation{Department of Astronomy, School of Science, The University of Tokyo, 7-3-1 Hongo, Bunkyo, Tokyo 113-0033, Japan}

\author{Takuma Izumi}
\affiliation{National Astronomical Observatory of Japan, 2-21-1 Osawa, Mitaka, Tokyo 181-8588, Japan}
\affiliation{Department of Astronomical Science, Graduate University for Advanced Studies (SOKENDAI), 2-21-1 Osawa, Mitaka, Tokyo 181-8588, Japan}

\author{Wanqiu He}
\affiliation{National Astronomical Observatory of Japan, 2-21-1 Osawa, Mitaka, Tokyo 181-8588, Japan}

\author{Masayuki Akiyama}
\affiliation{Astronomical Institute, Tohoku University, Aramaki, Aoba, Sendai 980-8578, Japan}

\author{Kohei Inayoshi}
\affiliation{Kavli Institute for Astronomy and Astrophysics, Peking
University, Beijing 100871, China}

\author{Yoshiki Matsuoka}
\affiliation{Research Center for Space and Cosmic Evolution, Ehime University, 2-5 Bunkyo-cho, Matsuyama, Ehime 790-8577, Japan}

\author{Masafusa Onoue}
\affiliation{Kavli Institute for Astronomy and Astrophysics, Peking
University, Beijing 100871, China}
\affiliation{Kavli Institute for the Physics and Mathematics of the Universe (WPI), The University of Tokyo, Kashiwa, Chiba 277-8583, Japan}

\author{Yoshiki Toba}
\affiliation{Department of Astronomy, Kyoto University, Kitashirakawa-Oiwake-cho, Sakyo-ku, Kyoto 606-8502, Japan}
\affiliation{Academia Sinica Institute of Astronomy and Astrophysics, 11F of Astronomy-Mathematics Building, AS/NTU, No.1, Section 4, Roosevelt Road, Taipei 10617, Taiwan}
\affiliation{Research Center for Space and Cosmic Evolution, Ehime University, 2-5 Bunkyo-cho, Matsuyama, Ehime 790-8577, Japan}

\begin{abstract}

\noindent The correlation between the mass of supermassive black holes (SMBHs; \mbh) and their host galaxies (\m) in the reionization epoch provides valuable constraints on their early growth. High-redshift quasars typically have a \mbh/\m~ratio significantly elevated in comparison to the local value. However, the degree to which this apparent offset is driven by observational biases is unclear for the most distant quasars.
To address this issue, we model the sample selection and measurement biases for a compilation of 20 quasars at $z\sim6$ with host properties based on ALMA observations. 
We find that the observed distribution of quasars in the \rel~plane can be reproduced by assuming that the underlying SMBH population at $z\sim6$ follows the relationship in the local universe. However, a positive or even a negative evolution in \mbh/\m~can also explain the data, depending on whether the intrinsic scatter evolves and the strength of various systematic uncertainties.  To break these degeneracies, an improvement in the accuracy of mass measurements and an expansion of the current sample to lower \mbh~limits are needed.
Furthermore, assuming a radiative efficiency of 0.1 and quasar duty cycles estimated from the active SMBH fraction, significant outliers in \mbh/\m~tend to move toward the local relation given their instantaneous BH growth rate and star formation rate. This may provide evidence for a self-regulated SMBH--galaxy coevolution scenario that is  in place at $z\sim6$, with AGN feedback being a possible driver.
\end{abstract}


\section{Introduction}
\label{sec:intro}
Active galactic nuclei (AGN) or quasars, powered by mass accretion onto SMBHs, produce an enormous amount of energy that has been long-speculated to have profound impacts on galaxy evolution \citep[e.g.,][]{King2015}. In the local universe, the mass of SMBHs appears to be closely connected to the bulge properties (e.g., bulge mass \mbulge, stellar velocity dispersion $\sigma_\star$), which inspired the concept of ``coevolution'' in studies of SMBH and galaxy evolution \citep[e.g.,][]{Kormendy2013}. 

High-redshift studies have mainly focused on the \rel~relation, with mounting evidence showing that its evolution in massive systems is not significant since $z\sim2$ \citep[e.g.,][]{Jahnke2009, Schramm2013, Sun2015, Ding2020a, Li2021}. In particular, its intrinsic scatter appears similar to the local value \citep[e.g.,][]{Ding2020a, Li2021}.  These results suggest that a physical coupling between SMBHs and galaxies (e.g., through AGN feedback) is likely at work to keep \mbh/\m~relatively constant. To decipher how the relationship was first established in the early universe, a key strategy would be to measure the \rel~relation in the reionization era ($z>6$) where we are able to probe the first generation of accreting SMBHs \citep[e.g.,][]{Inayoshi2020}.

Many of the $z\sim6$ quasars discovered so far are powered by extremely massive BHs with $\mbh \sim 10^9\ \msun$ \citep[e.g.,][]{Fan2000, Shen2019}, and are actively forming stars with star formation rates (SFR) $\rm \sim 100-1000\ \msun\ yr^{-1}$ \citep[e.g.,][]{Wang2013}. Their \mbh/\m~(where \m~is approximated by the dynamical mass \mdy~measured from gas kinematics using ALMA) appears to be significantly offset from the local value by up to $2.0$~dex, suggesting that the growth of SMBHs substantially precedes their hosts \citep[e.g.,][]{Neeleman2021}. However, these quasars are biased tracers of the underlying SMBH population, since only the most luminous quasars powered by the most massive BHs can be detected in shallow surveys \citep[e.g.,][]{Lauer2007, Volonteri2011, Schulze2014}. Lower-luminosity quasars detected in deeper surveys (e.g., the SHELLQs survey; \citealt{Matsuoka2016}) lie closer to the local relation thus confirming this bias \citep[e.g.,][]{ Izumi2019, Izumi2021}. 

Moreover, the mass measurements at high redshifts suffer from significant uncertainties with possibly systematic biases. For instance, \mbh~of a flux-limited quasar sample might be statistically overestimated by the single-epoch virial estimator (e.g., \citealt{Vestergaard2009}; hereafter VO09) because of uncorrelated variation between AGN luminosity and broad line width (especially for \mgii~and \civ). This gives rise to a luminosity-dependent virial BH mass bias (hereafter the SE bias) as detailed in \cite{Shen2013}. In addition, the hosts of $z\sim6$ quasars are found to be gas-rich \citep[e.g.,][]{Decarli2022}, thus approximating \m~by \mdy~is likely an overestimate.

In this Letter, we model the selection and measurement biases for a $z\sim6$ quasar sample compiled in the literature in order to reveal the underlying connection between SMBH and galaxy growth in the early universe. This work assumes a cosmological model with $\Omega_\Lambda$ = 0.7, $\Omega_{\rm m}$ = 0.3, and $H_0 = 70\rm\ km\ s^{-1}\ Mpc^{-1}$. The stellar mass and SFR are based on the \cite{Chabrier2003} initial mass function.

\begin{figure*}[!htbp]
\includegraphics[width=\linewidth]{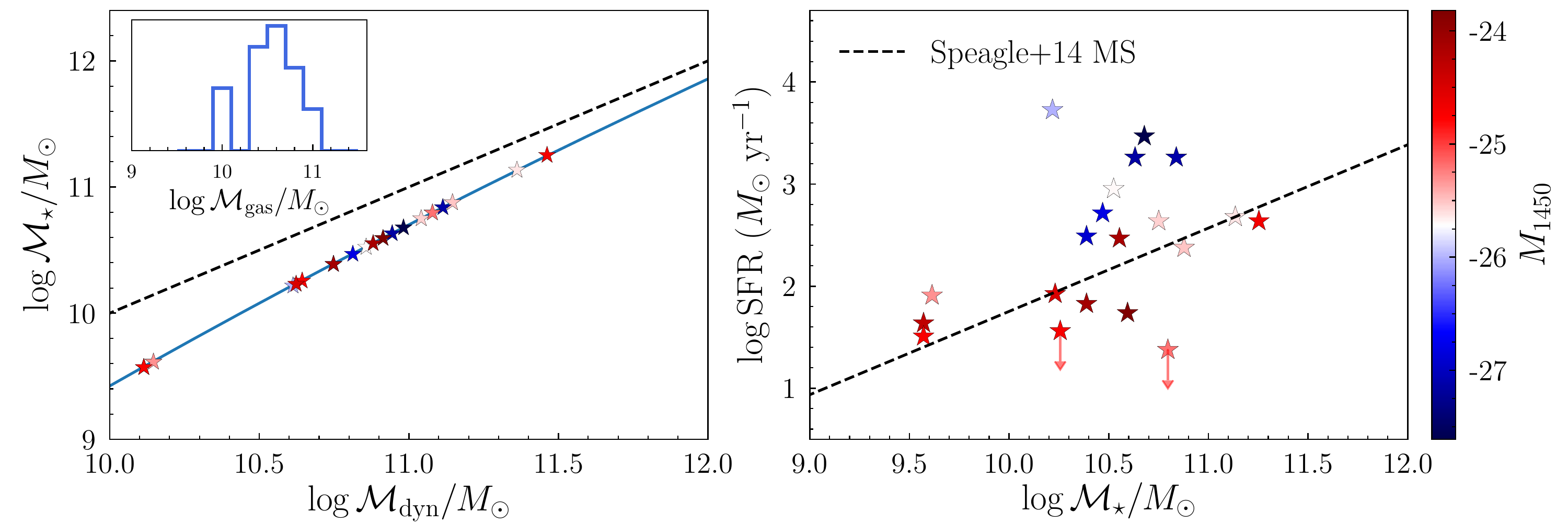}
\caption{Left: Estimating stellar masses of $z\sim6$ quasars by subtracting the expected molecular gas masses (inset histogram) from their dynamical masses. Quasars are plotted as stars color-coded by their \muv. The blue curve shows the \m--\mdy~relation derived from the \mgas/\m~vs. \m~relation given by \cite{Tacconi2018}. The black dashed line shows the one-to-one relation. Right: The SFR--\m~plane of our quasars compared to the \cite{Speagle2014} main sequence relation at $z\sim6$.  Two objects with upper limits in SFR are marked by red arrows.}
\label{fig:mstellar}
\end{figure*}

\section{Sample}
\label{sec:sample}

We adopt the $z\sim6$ quasar sample compiled in \cite{Izumi2021} as our parent sample. It contains 46 quasars with \mbh, \mdy, quasar luminosity ($L_{3000}$ and \muv), and infrared (IR) luminosity measurements from the literature. The \mbh~of 22/46 objects are derived from the virial estimator using the VO09 calibration for the \mgii~line as
\begin{equation}
\begin{split}
{\rm log} \Bigg(\frac{\mbh}{\msun} \Bigg) = 6.86 + 0.5\, {\rm log}\,\Bigg(\frac{\lambda L_{\lambda3000}}{\,10^{44}\, \ergs} \Bigg)\\ +  2\,{\rm log}\,\Bigg(\frac{\nuw}{\rm km\ s^{-1}}\Bigg),
\label{eq:mbh}
\end{split}
\end{equation}
while Eddington-limited accretion is assumed to estimate the mass for the remaining 24 objects.

The total IR luminosity ($L_{\rm TIR}$) is derived by fitting the 1.2~mm ALMA continuum with an optically thin graybody spectrum assuming a dust temperature of 47~K and a dust spectral emissivity index of 1.6 that have been regularly adopted for $z\sim6$ quasars, then extrapolating to the total IR ($8-1000\ \mu \rm m$) range; although the actual dust temperature could vary from source to source \citep[e.g.,][]{Venemans2016}. The SFR is derived using $3.88 \times 10^{-44}\, L_{\rm TIR}$  \citep{Murphy2011}, assuming that the cold interstellar medium is predominantly heated by star formation.

The dynamical masses \mdy~of these quasars are derived through gas kinematics using the \cii~line.
The standard rotating thin disk approximation is assumed for all quasars except two (given as upper limits). In this work, we only consider the 20 objects whose \mbh~and \mdy~are derived from the \mgii~line and the rotating thin disk assumption, respectively, to ensure relatively reliable mass measurements. However, we caution that the derived \mdy~is highly sensitive to the assumptions made on galaxy geometry and inclination angle \citep[e.g.,][]{Wang2013}.

Ignoring the possibly small contribution of dark matter within the \cii~emitting region \cite[e.g.,][]{Genzel2017}, we estimate the \m~of these quasars by subtracting the expected molecular gas mass (\mgas) from their total \mdy, assuming that quasar hosts have similar gas content as star forming galaxies \citep[e.g.,][]{Molina2021}.
We adopt the typical \mgas/\m~ratio ($\mu_{\rm gas}$) of $z\sim6$ galaxies given by \cite{Tacconi2018}:
\begin{align*}
    {\rm log}\,\mu_{\rm gas} = & \ 0.12 - 3.62 \times ({\rm\,log}\,(1+z) - 0.66)^2 \\ 
    & - 0.35\times (\logm-10.7), 
\end{align*}
where we adopted their $\beta=2$ result with the \cite{Speagle2014} star formation main sequence (MS) and assumed  $\delta{\rm MS}=0$. We then use this relationship to estimate the typical \mgas~at a given \m~and derive the correlation between \m~and \mdy, where \mdy~is approximated by $\m+\mgas$. The resulting \mgas~and \m~of our quasars at their respective \mdy~are shown in Figure \ref{fig:mstellar} (left panel). The gas mass is distributed between $\sim10^{10}-10^{11}\ \msun$, which is consistent with recent direct measurements in $z\sim6$ quasars \citep[e.g.,][]{Decarli2022}. The derived \m~is typically $0.2-0.5$ dex smaller than \mdy. We also show the distribution of our quasars in the SFR -- \m~plane in Figure \ref{fig:mstellar} (right panel). As can be seen, most quasars are located near the \cite{Speagle2014} MS.

\section{Evolution of the \rel~Relation}
\label{sec:connection}

\subsection{The Observed \rel~Relation}
\label{subsec:obs}

Figure \ref{fig:obs_mbh} shows the $z\sim6$ quasars in the \rel~plane compared to the local \bulgerel~relation. We adopt the local relation given by \cite{Haring2004} (HR04) to be self-consistent with the VO09 virial estimator (see Section 6.2 in \citealt{Ding2020a} for the discussion on the choice of the local baseline). The local sample consists of massive ellipticals and bulge-dominated S0 galaxies, thus we adopt $\mbulge \approx \m$ in the following analyses. This figure confirms that quasars at $z\sim6$ typically lie above the local relation, and the offset at a given stellar mass generally increases with quasar luminosity.

\subsection{Estimating Expected Biases}
\label{subsec:bias}

As introduced in Section \ref{sec:intro}, the current $z\sim6$ quasar sample suffers from strong selection biases.  Following \cite{Li2021}, we perform a simple Monte Carlo simulation to build a mock AGN sample that mimics the observational biases to account for such an effect in order to reveal the underlying connection between SMBHs and their host galaxies \cite[e.g.,][]{Schulze2014, Volonteri2016}. In the following, we briefly introduce our simulation method. The details of each step and the choice of model parameters are described in Appendix \ref{sec:appendix}.

Our simulation starts with the galaxy stellar mass function (SMF) at $z\sim6$ given by \cite{Grazian2015} and the \rel~relation to generate a sample of mock galaxies and SMBHs. The \rel~relation is assumed to have a gaussian intrinsic scatter (\sigmamu) with a mean that evolves as \begin{equation}
\deltalogmbh \equiv {\rm log}\, (q_z /q_{z=0}) = \gamma\, {\rm log}\,(1+z),
\label{eq:evolution}
\end{equation}
where $q$ is the mass ratio \mbh/\m.
Assuming that type 1 AGNs follow the same \rel~relation as the underlying galaxy population (see \citealt{Schulze2014} and \citealt{Li2021}), we determine the bolometric luminosity by using \mbh~and adopting an intrinsic Eddington ratio (\edd) distribution function (ERDF) of type 1 AGNs. We use the ERDF at $z=4.75$ given by \cite{Kelly2013} who jointly constrained the intrinsic ERDF and the active BH mass function (BHMF) based on uniformly-selected SDSS quasars with the sample incompleteness being carefully corrected in a Bayesian framework.

We then derive a virial \mbh~for each mock AGN using the VO09 virial estimator based on its true \mbh, luminosity, an assumed FWHM distribution (in a log-normal form; \citealt{Shen2013}), and a parameter $\beta$ ($0 \leqslant \beta \leqslant1$) that describes the fraction of correlated response of line width to the variation of luminosity. The value of $\beta$ is poorly constrained at present. We adopt $\beta=0.6$ in this work, while $\beta \neq 1$ will give rise to the SE bias \citep{Shen2013}. The resulting virial \mbh~has a 0.4~dex scatter relative to the true \mbh~and it tends to overestimate the true \mbh~if $L>\overline{L}$ (and vice versa), where $L$ is the luminosity of an AGN with a true BH mass of $\mathcal{M}_{\rm BH, true}$, and $\overline{L}$ is the average luminosity of all AGNs at the same $\mathcal{M}_{\rm BH, true}$ (see Appendix \ref{sec:appendix} for details).
In addition, we add a random gaussian error with a dispersion of $0.5$~dex to each true \m~to reflect the large uncertainties of estimating \m~from \mdy. 

In our framework, under the assumption of no evolution in the \rel~relation (i.e., $\gm = 0.0$, $\sigmamu = 0.3$) of the underlying SMBH population, one can estimate the expected bias (i.e., a positive \deltalogmbh~relative to the local relation) caused by the sample selection function by applying the same selection criteria of observations to mock AGNs. However, it is infeasible to define a selection function as our sample is a mixture of $z\sim6$ quasars from various surveys with additional requirements of having near-IR spectroscopic and ALMA follow up to measure \mbh~and \mdy. Therefore, we assume a simplified scenario for which all  selection biases come from the ``effective'' magnitude limit of different surveys, where effective means that these quasars are the relatively luminous ones selected from their parent samples for follow up observations. We simulate this selection function by producing randomly drawn samples of mock quasars that are matched to the observed \muv~distribution (hereafter the Mock-Q sample).

\begin{figure}
\includegraphics[width=\linewidth]{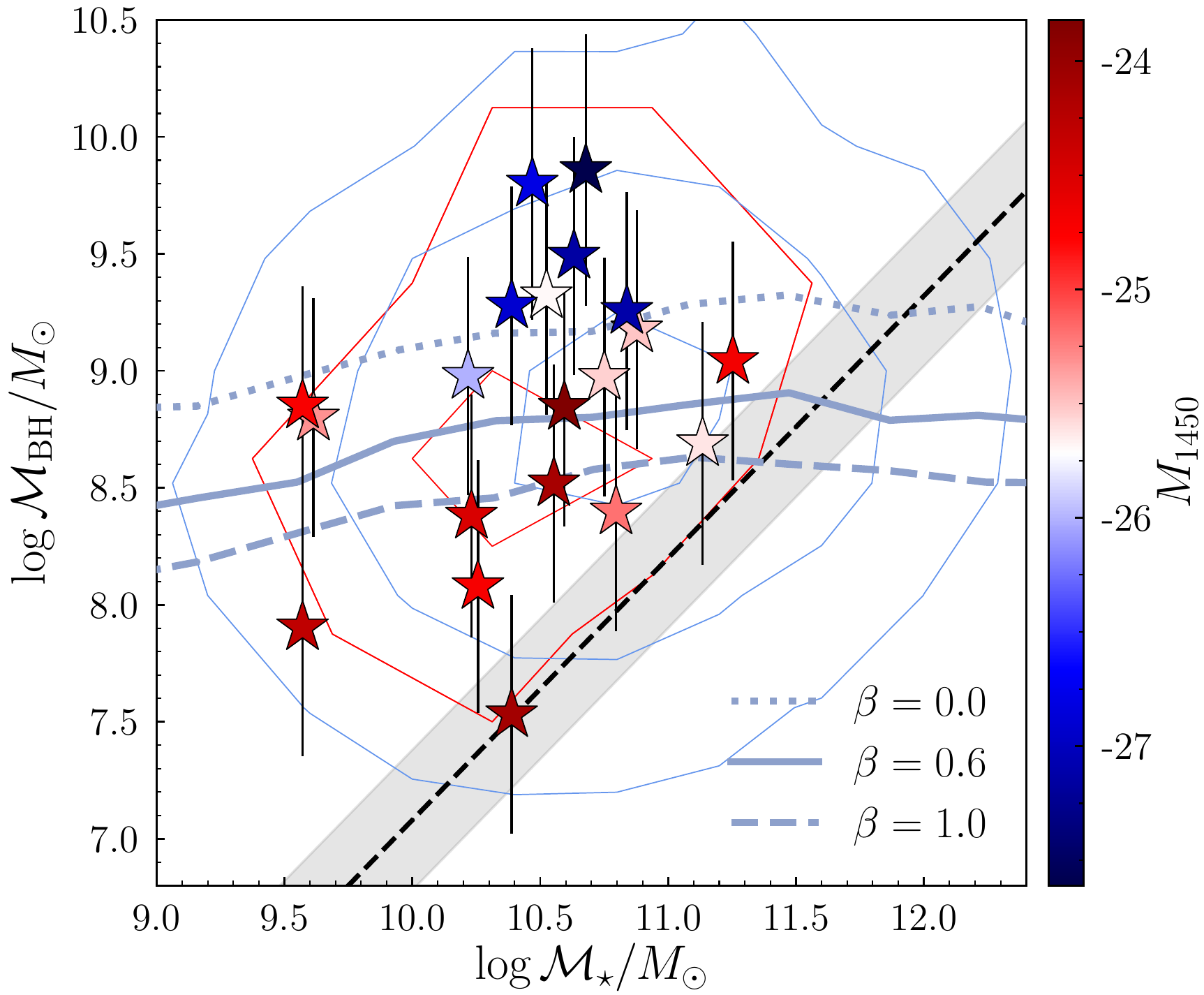}

\caption{The \rel~distribution at $z\sim6$. The individual points and the red contours ($1-2\,\sigma$ levels) show the observed quasars color-coded by their \muv. The blue contours ($1-3\,\sigma$ levels) show the mock AGN sample which mimics the selection function and measurement uncertainties of observations (i.e., Mock-Q) assuming $\beta=0.6$. The average \rel~relations of the mock AGN samples (incorporating observational biases) using $\beta=0.0$, 0.6, and 1.0 are plotted as dotted, solid, and dashed blue curves, respectively. The local HR04 relation and its scatter are shown as a black dashed line and a gray shaded region.}
\label{fig:obs_mbh}
\end{figure}

The distribution of the Mock-Q sample in the \rel~plane is plotted as blue contours in Figure \ref{fig:obs_mbh}. Their virial \mbh~tends to overestimate the true \mbh~by $\sim0.25$ dex (see Figure \ref{fig:params}e in the Appendix). 
Given the magnitude limit and the large uncertainties being added to both masses, the distribution is strongly modulated compared to the originally assumed HR04 relation. 
In Figure \ref{fig:obs_mbh} we also show the average virial \mbh~in bins of \m~for the Mock-Q sample as a blue solid curve. It represents the expected offset caused by selection effects and measurement uncertainties. To rephrase, any offset and large scatter in the observed \rel~relation (red contours in Figure \ref{fig:obs_mbh}) that follows the blue curve and contours could be considered as lacking significant evolution in the mass relation of the underlying SMBH population, which appears to be true for our quasar sample. Note that the small offset between the observed quasars and the model predictions could be due to the different methods used to derive stellar masses in \cite{Grazian2015} and this work (SED fitting vs. $\mdy-\mgas$ where the latter may underestimate the total stellar mass; see Section \ref{subsec:intrinsic}). 

We also show the impact of varying $\beta$ in Figure \ref{fig:obs_mbh}. For reference, the choice of an extreme value (unlikely to be true; see Appendix \ref{sec:appendix} for details) for $\beta$ will cause systematic
shifts of the average \rel~relation for the Mock-Q
sample by about $-0.2$~dex and +0.4 dex for the respective case of $\beta=1.0$ (i.e., no SE bias) and $\beta=0.0$ (i.e.,
the maximum SE bias for which the BH mass tends to
be overestimated by $\sim0.65$~dex). In both cases, the expected positions of $z\sim6$ quasars (i.e., the dashed and dotted curves) are offset from the actual observed ones, thus a more positively evolving (for $\beta=1.0$) or a more negatively evolving (for $\beta=0.0$) \rel~relation is required to explain the observations.

\subsection{Constraining the Intrinsic \rel~Relation}
\label{subsec:intrinsic}

\begin{figure}
\includegraphics[width=\linewidth]{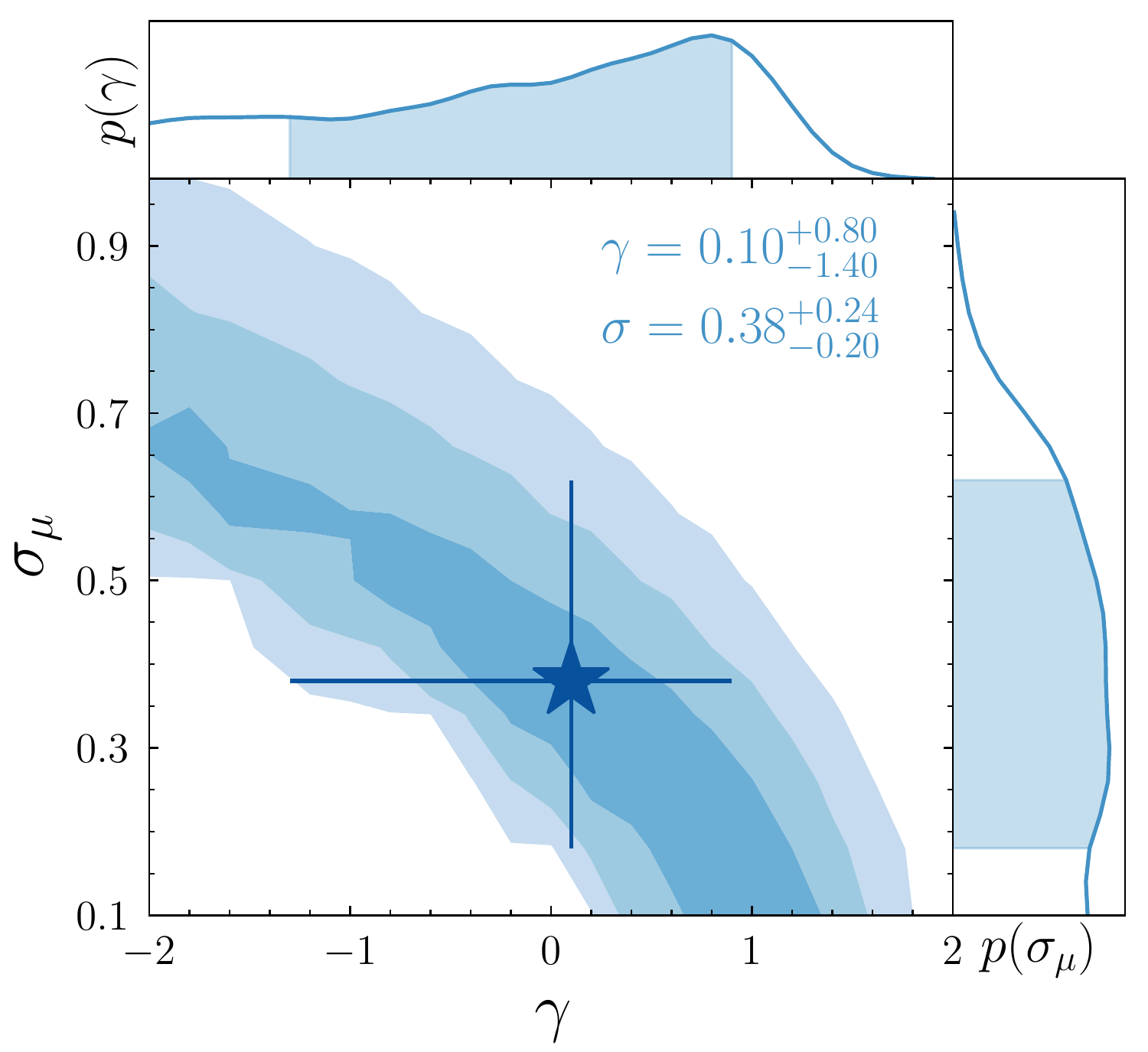}
\caption{Constraining the evolution factor \gm~and the intrinsic scatter \sigmamu~based on a specific model ($\beta=0.6$). The contours represent the $1-3\,\sigma$ confidence regions of the posterior probability distribution. The histograms are the marginalized posterior distributions. The star indicates the posterior median and its $1\,\sigma$ error.}
\label{fig:joint}
\end{figure}

While the observed offset can be reproduced by an unevolving \rel~relation, evolutionary models cannot be ruled out. Therefore, we determine the constraints on the intrinsic evolution by generating mock AGNs with a range of \gm~and \sigmamu~(assuming $\beta=0.6$) and comparing the resulting \rel~relation (incorporating observational biases) with the observed one to derive the likelihood (see Section 5 in \citealt{Li2021} for details).

The posterior distributions of \gm~and \sigmamu~assuming a bounded flat prior ($-2.0 <\gm< 2.0$, $\sigmamu>0.1$) are shown in Figure \ref{fig:joint}. There is a strong degeneracy between \gm~and \sigmamu: either a positive evolution (i.e, \mbh/\m~increases with redshift) with a small \sigmamu, or  a negative evolution with a large \sigmamu~can reproduce the large apparent offsets. 
The peak of the posterior distribution is slightly skewed towards a positive evolution with a small scatter, as the  sample only probes overly massive quasars that are clustered at the top left of the local relation. The best-fit values based on the 16th, 50th, and the 84th percentiles of the marginalized posterior distributions are $\gm=0.10_{-1.40}^{+0.80}$ and $\sigmamu=0.38_{-0.20}^{+0.24}$, which is consistent with an unevolving \mbh/\m~out to $z\sim6$ \citep[e.g.,][]{Volonteri2011, Schulze2014}. 

However, given various systematic uncertainties involved in mass measurements and model assumptions, it is improper to interpret the complex evolution with a single number based on a certain model. For instance, the simplified thin-disk approximation will yield an underestimated \mdy~(thus \m) if the galaxy contains a dispersion-dominated component \citep[e.g.,][]{Pensabene2020}. In addition, while we intended to use the total \m~of these quasars in our analysis which has been shown to correlate better with \mbh~than using \mbulge~at high redshifts \citep[e.g.,][]{Jahnke2009, Schramm2013, Li2021}, the \cii~emission line mainly traces the inner galaxy region but not the entire galaxy \citep[e.g.,][]{Venemans2017}. These effects, together with the uncertain gas fraction and the choice of $\beta$, can  induce systematic shifts of the mass measurements and affect the posterior distributions. Moreover, the bias estimates also depend on the assumed underlying distribution functions (e.g., the ERDF), which are not well-constrained at $z\sim6$ (see Section 6.3 in \citealt{Li2021} for a detailed discussion). Therefore, it is clear that the current dataset is insufficient to robustly constrain the intrinsic \rel~relation of the underlying SMBH population at $z\sim6$. 

\subsection{Predicted Evolution in the \rel~Plane}

\begin{figure}
\includegraphics[width=\linewidth]{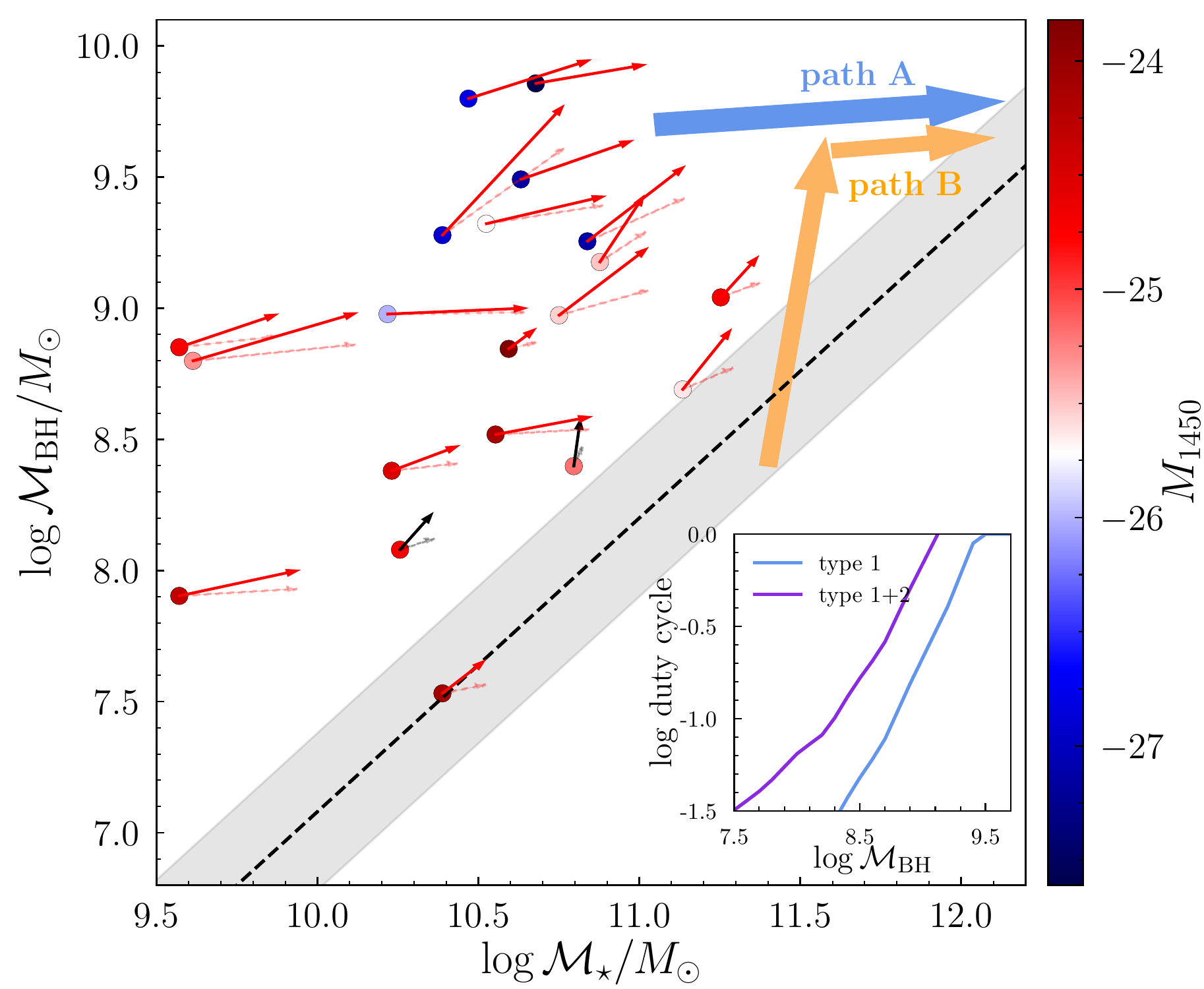}
\includegraphics[width=\linewidth]{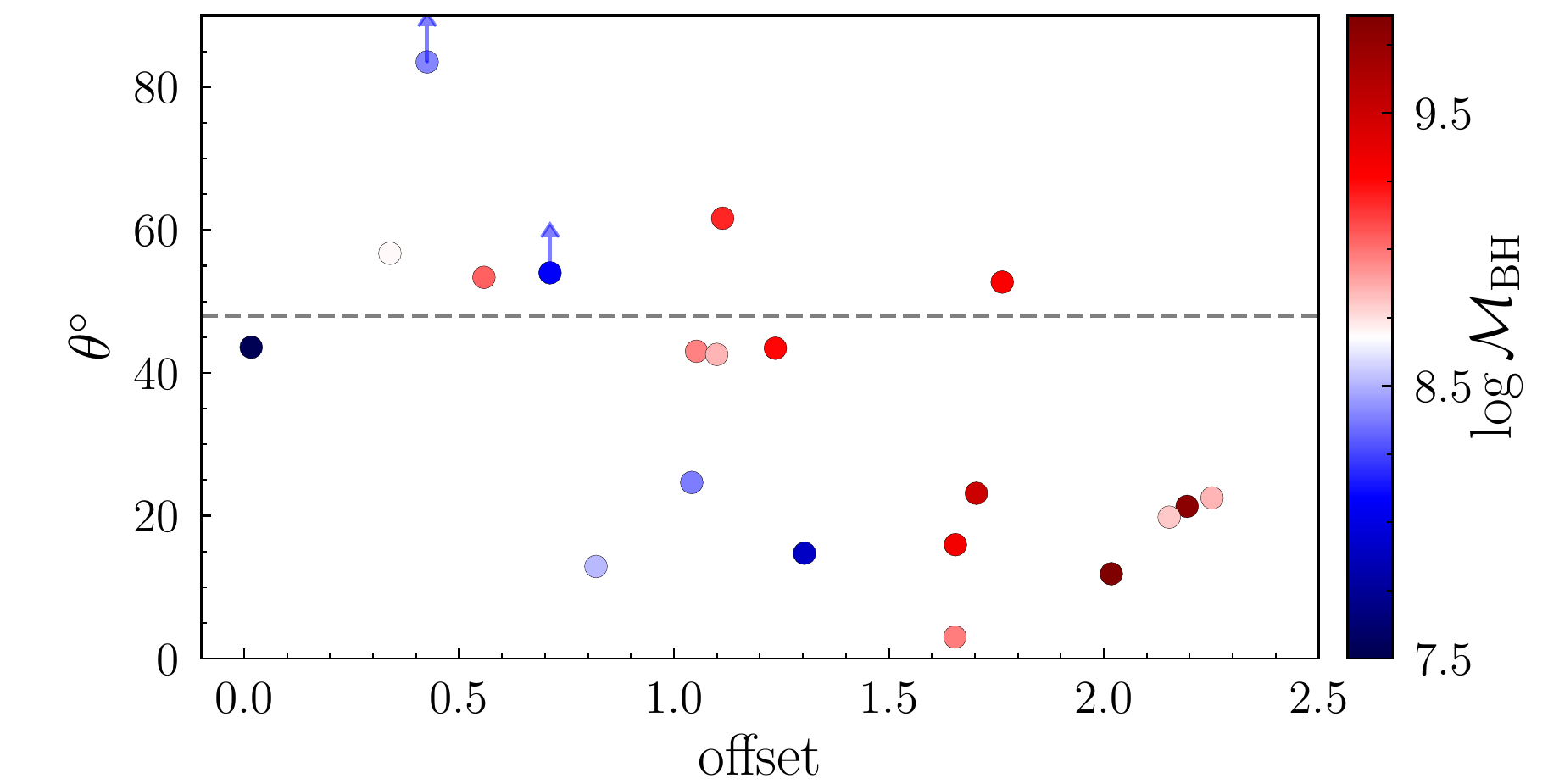}
\caption{Top: predicted evolution of the growth of SMBHs and their host galaxies over the next min($t_{\rm del}$, 100) Myr. The dashed arrows represent the evolution vectors assuming the AGN duty cycle derived from the type 1 BHMF, while the solid arrows show the evolution vectors after correcting the duty cycle for type 2 AGNs. Two objects with upper limits in SFR are shown as black (gray) arrows. 
The adopted duty cycle -- \mbh~relations are shown in the inset. The blue and orange arrows show the possible evolution pathways for overly massive quasars. The black dashed line and the shaded region represent the local HR04 relation and its scatter. Bottom: the intersection angle as a function of offset. Each quasar is color-coded by their BH masses. Two objects with upper limits in SFR are marked by blue arrows. The dashed line at $\theta=48^\circ$ represents the slope of the local relation.}
\label{fig:flow}
\end{figure}

Although the intrinsic evolution is unclear, overly massive quasars do exist. It is inevitable that their vigorous BH growth needs to be inhibited at some point in order to avoid unreasonably large \mbh~and to prevent them from further deviating from the local relation.
The subsequent locations of these quasars in the \rel~plane could be assessed by their instantaneous BH growth rate (BHGR) and SFR \citep[e.g.,][]{Venemans2016}. 
The BHGR can be estimated as 
\begin{equation}
    \dot{\mbh} = \frac{(1-\eta)\,\lbol}{\eta c^2},
\end{equation}
where $c$ is the speed of light and $\eta=0.1$ is the assumed radiative efficiency. The adopted constant $\eta$ is suitable for our moderately accreting SMBHs that span $0.15 \lesssim \edd \lesssim 3.0$ as expected from  standard thin disk theory \citep{Shakura1973} or slim disk theory \citep{Abramowicz1988} for mildly super-Eddington quasars \citep[e.g.,][]{Inayoshi2019disk, Inayoshi2020}. We assume that these $z\sim6$ quasars can continue to form stars at their current SFR for a period ($\Delta t$). At the same time, the SMBHs keep accreting at the measured BHGR for a fraction of this time (i.e., the AGN duty cycle). The challenge is to constrain how long a quasar can sustain its high growth rate with its own feedback \citep[e.g.,][]{Valentini2021}; recent observations report a short quasar lifetime at $z\sim6$ ($\sim10^6$~yr on average; \citealt{Eilers2021}). We estimate the \mbh-dependent duty cycle from the active SMBH fraction (see the inset in Figure \ref{fig:flow}), which is derived from the ratio between the BHMF of type 1 AGNs at $z\sim4.75$ \citep{Kelly2013} to the total BHMF scaled from the SMF \citep{Grazian2015} at the same redshift and assuming a \rel~relation with $\gm=0.1$ and $\sigmamu=0.38$ (Section \ref{subsec:intrinsic}). This method likely yields an upper limit on the BH growth time at the quasar accretion rate as most active BHs at a given \mbh~are of much lower luminosities.
We adopt $\Delta t$ as the minimum value between the gas depletion timescale ($t_{\rm del}=\mathcal{M}_{\rm gas}$/SFR; $\sim10-1000$~Myr) and 100~Myr (arbitrary value chosen to better visualize the evolution trend, which is shorter than most $t_{\rm del}$). The direction of \mbh~and \m~during this period are illustrated by the dashed arrows in the top panel of Figure \ref{fig:flow}, with the caveats that our estimation is a simplification of the complex physical processes (e.g., AGN feedback, gas accretion from the environment, merger) that could happen over the next $\sim100$~Myr and the duty cycles for individual quasars are prone to significant uncertainties that are impossible to accurately constrain at present.

Interestingly, the predicted evolution exhibits a flow pattern, where quasars that are significant outliers in \mbh/\m~tend to converge to the local relation as previously seen at $z<2$ \citep[e.g.,][]{Merloni2010, Sun2015}.  We also show the evolution vectors (see the solid arrows) after correcting the active fraction for type 2 AGNs (see the inset in Figure \ref{fig:flow}) assuming the luminosity-dependent obscured fraction ($\sim60-80\%$ at $z\sim6$) given by \cite{Vito2018} to account for the possible underestimate of the BH growth time that is not captured by the BHMF of type 1s. Still, the general trend remains. 
The converging pattern also holds if we adopt the peak value of $\gamma$ and \sigmamu~when estimating the total BHMF which results in lower AGN duty cycles. 
However, the SFR derived from the single-band ALMA photometry may be overestimated since quasars could also contribute to rest-frame $\rm \sim 158\ \mu m$ emissions \citep[e.g.,][]{McKinney2021}. Taking this effect into account will make the converging trend weaker or even disappear if the SFR is overestimated by a factor of $\gtrsim2$. Multi-band ALMA photometry are thus crucial to accurately constrain the dust temperature and subtract the quasar contamination when deriving the SFR. 

In the bottom panel of Figure \ref{fig:flow} we plot the intersection angle $\theta$ between the evolution vectors derived from the type 1+2 duty cycle and the local relation as a function of offset (color-coded by \mbh) where a decreasing trend is evident. There are 12 of 14 quasars with an offset larger than 1.0 dex have $\theta < 48^\circ$, where $\theta=48^\circ$ corresponds to the slope of the local relation. The median intersection angle at ${\rm offset} >1.0$ dex is $\approx 23^\circ$, which is significantly smaller than that at ${\rm offset}<1.0$~dex ($\theta \approx 54^\circ$). It can also be seen that at similar BH masses, $\theta$ is smaller for quasars with larger offsets, thus the decreasing trend is not driven by less massive BHs with smaller offsets and shorter duty cycles. A natural explanation of the flow pattern and the decreasing trend could be AGN feedback \citep[e.g.,][]{Valentini2021}, which suppresses the growth of SMBHs once they deviate significantly from the local relation. 

The converging pattern for the most luminous and massive BHs may suggest that they have experienced rapid acccretion episodes during seeding epochs and remain being overly massive until reaching the local relation (i.e., path A in Figure \ref{fig:flow}) as shown by recent numerical simulations \citep[e.g.,][]{Inayoshi2022}. 
However, we cannot rule out the possibility that their progenitors are low-mass BHs moving upwards at similar stellar masses (i.e., path B in Figure \ref{fig:flow}). 
Such low-\mbh~objects are undersampled in current surveys due to the detection limit, and their vigorous BH accretion may occur rapidly in a highly obscured or/and a radiative inefficient mode which further reduces their apparent luminosity \citep[e.g.,][]{Trebitsch2019, Davies2019}. Therefore, it is essential to study the growth of low mass systems and obscured quasars that have recently been discovered at $z\sim6$ \citep[e.g.,][]{Onoue2021}.

\section{Concluding Remarks}
\label{sec:discussion}
The $z\sim6$ quasars typically have \mbh/\m~significantly larger than the local value. However, strong selection biases and significant measurement uncertainties severely limit the interpretation of the data. In this work, we account for these factors and demonstrate that the large apparent offsets and observed scatter could be reproduced by assuming that the underlying SMBH population at $z\sim6$ follows the local \rel~relation (Figure \ref{fig:obs_mbh}). However, a positive or even a negative evolution can also explain the data, depending on the evolution of the intrinsic scatter and various systematic uncertainties (Figure \ref{fig:joint}). It is thus crucial to emphasize that the evolution of the offset cannot be properly assessed without considering the scatter (see \citealt{Li2021} for a similar issue at $z<1$). Interestingly, quasars that are significant outliers in \mbh/\m~tend to have evolution vectors pointing toward the local relation (Figure \ref{fig:flow}). This may provide  evidence that a self-regulated SMBH-galaxy coevolution scenario is already in place at $z\sim6$, possibly driven by AGN feedback, although a robust conclusion can only be achieved with future observations that can accurately constrain the SFR (currently estimated from a single-band ALMA photometry assuming that the cold interstellar medium is mainly heated by star formation) and duty cycle for these quasars.

To break the degeneracy, expanding the current sample in both number statistics and to lower \mbh~limits are imperative \citep[e.g.,][]{Habouzit2022}. This is expected to be achieved by the ongoing SHELLQs survey \citep[e.g.,][]{Matsuoka2016} and the forthcoming surveys by the Vera C. Rubin Observatory and Euclid, which will offer promisingly large and less-biased quasar samples with a more uniform selection function. It is also crucial to reduce the uncertainties (especially systematic effects) in the mass measurements. The James Webb Space Telescope will enable us to measure \mbh~using the more reliable \hb~line, and makes it possible to probe the stellar emissions of $z\sim6$ quasars in the rest-frame optical bands thus allowing direct measurements of their \m~\citep[e.g.,][]{Marshall2021}. In the meantime, extending the current reverberation-mapped AGN sample to a wider parameter space is also important to validate and improve the virial \mbh~estimator, which will be achieved by the ongoing SDSS-V Black Hole Mapper survey. These efforts, together with a deeper understanding of the AGN accretion process (e.g., radiative efficiency, duty cycle) will allow us to better assess the connection between SMBH and galaxy growth in the reionization era.

\acknowledgments
We thank the referee for valuable suggestions that helped to improve the manuscript.
J.Y.L. acknowledges support from the National Natural Science Foundation of China (12025303, 11890693). J.D.S. is supported by the JSPS KAKENHI Grant Number JP18H01251, and the World Premier International Research Center Initiative (WPI Initiative), MEXT, Japan. K.I. acknowledges support from the National Natural Science Foundation of China (12073003, 12003003, 11721303, 11991052, 11950410493), the National Key R\&D Program of China (2016YFA0400702), and the China Manned Space Project with NO. CMS-CSST-2021-A04 and CMS-CSST-2021-A06.

\begin{appendix}
\section{Details of Generating a Mock AGN Sample}
\label{sec:appendix}
\begin{figure*}
\includegraphics[width=\linewidth]{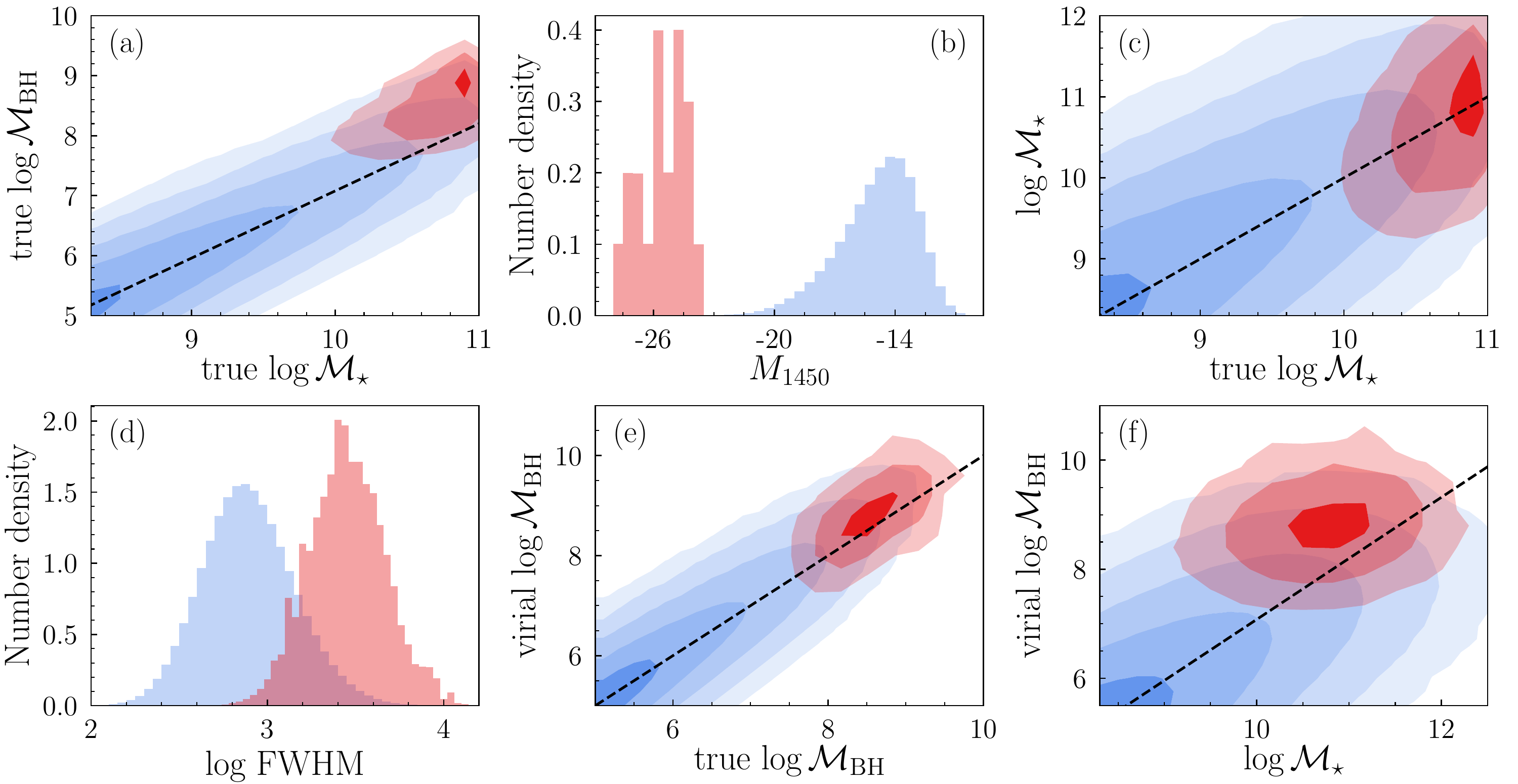}
\caption{Parameter distributions of the mock AGN sample. The full mock sample is shown in blue with the contours representing the $1-5\,\sigma$ levels. The mock sample matching in \muv~with the observed $z\sim6$ quasars is shown in red with the contours representing the $1-3\,\sigma$ levels.}
\label{fig:params}
\end{figure*}

Here we outline the steps in our simulation by assuming $\gm=0.0$ and $\sigmamu=0.3$ as an example. The simulation starts with the SMF at $z\sim6$ given by \cite{Grazian2015} to generate a sample of mock galaxies ranging from $8.0<\logm/\msun<11.0$. We assume that the \rel~relation at $z\sim6$ is the same as the local HR04 relation in terms of both mean and intrinsic scatter, and randomly assign each mock galaxy a {\it{true}} \mbh~based on the HR04 relation. The resulting \rel~distribution is shown in blue in Figure \ref{fig:params}a.

We convert \mbh~into bolometric luminosity by randomly sampling the intrinsic ERDF given by \cite{Kelly2013} at $z=4.75$ in the range of $-1.5<\logedd<0.5$ \citep[e.g.,][]{Shen2019, Onoue2019}. The bolometric luminosity is converted into rest-frame luminosity $L_{\rm 3000}$ and absolute magnitude $M_{\rm 1450}$ (Figure \ref{fig:params}b) assuming the bolometric corrections to be 5.15 and 4.4, respectively \citep{Richards2006}. After these steps, we have the full knowledge of the {\it{true}} distribution of mock AGNs in the $\m-\mbh-L$ plane. 

We then add realistic uncertainties to the mass terms to resemble observations. Given the large uncertainties on \mdy~and \mgas, we first add a random gaussian uncertainty with a standard deviation of 0.5~dex to each true \m~(Figure \ref{fig:params}c).  
We also derive a virial \mbh~for each mock AGN using their true \mbh, $L_{\rm 3000}$, and an assumed \mgii~FWHM distribution through Equation~\ref{eq:mbh}. In this step, we take into account the luminosity-dependent SE bias.
The SE bias originates from uncorrelated scatter between AGN luminosity and broad line width due to both the variability of an individual quasar and the object-by-object diversity in the broad-line region (BLR) properties at a fixed true \mbh~\citep{Shen2013}. We consider the following cases to represent different levels of the SE bias:
\begin{itemize}[leftmargin=*]

\item Case A:  virial \mbh~is an unbiased estimator of the true \mbh~regardless of AGN luminosity. This is done by assuming that the FWHM of \mgii~follows a log-normal distribution with the mean value determined by the {\it{true}} \mbh~and $L_{\rm 3000}$ for each mock AGN using Equation~\ref{eq:mbh}. We randomly sample the log-normal distribution with a dispersion of $\sigma_{\rm FWHM}$ to generate FWHM for each source. The sampled FWHM is then combined with $L_{\rm 3000}$ to derive virial \mbh. The dispersion $\sigma_{\rm FWHM}$ is chosen such that the resulting scatter of virial \mbh~to true \mbh~is 0.4~dex. By doing so, the variation of luminosity (relative to the mean) at a fixed true \mbh~is compensated by the concordant variation in FWHM. 

\item Case B: Only part of the variation in luminosity can be compensated by line width.  To simulate such a situation, for each mock AGN, we derive the difference ($\Delta L$) between its luminosity ($L$) and the mean luminosity ($\overline{L}$) for all mock AGNs of the same true \mbh. We assume that a fraction ($\beta$) of $\Delta L$ can be compensated by the line width, by using $\overline{L}$+$\beta \Delta L$ to determine the mean of the log-normal FWHM distribution for each source.
In this case, the higher (lower)-than-the-mean luminosity can only be partly compensated by the lower (higher)-than-the-mean line width, thus the resulting virial \mbh~tends to overestimate (underestimate) the true \mbh~except at  $L=\overline{L}$.
\end{itemize}

It is currently unclear how strong $\beta$ is. The non-breathing effect of \mgii~(i.e., the broad line width does not respond to the continuum variability in individual quasar) suggests that $\beta$ is not one \citep[e.g.,][]{Yang2020}. However, despite the lack of a BLR size $(R)-L$ relation for individual quasars (in case of \mgii), a global $R-L$ relation for a population of quasars spanning a broad range in BH masses and luminosities may still exist \citep[e.g.,][]{Homayouni2020}. This justifies the foundation of using the \mgii~line as a single-epoch virial estimator, thus $\beta$ is not likely to be zero. We adopt a relatively high response fraction ($\beta=0.6$) as our fiducial model. This assumption yields $\Delta {\rm log}\,\nuw \propto -0.15 \Delta L_{\rm 3000}$, as expected if the slope of the $R-L$ relation for \mgii~is $\sim 0.3$ \citep[e.g.,][]{Homayouni2020}. The resulting FWHM distribution and the virial \mbh~vs. true \mbh~relation are shown in Figures \ref{fig:params}d and \ref{fig:params}e, respectively.  

With the aforementioned steps, we have generated \m~and virial \mbh~for each mock AGN with realistic uncertainties (Figure \ref{fig:params}f). 
In Figure \ref{fig:params} we show the distributions of a mock AGN sample matching in \muv~with the observed $z\sim6$ quasars in red. The originally assumed underlying distributions are strongly modulated by the magnitude limit and the large uncertainties on both masses. In particular, the virial \mbh~tends to overestimate the true \mbh~by $\sim0.25$ dex for this specific mock sample (Figure \ref{fig:params}e), and the virial \mbh~vs. \m~relation is significantly offset from the HR04 relation (Figure \ref{fig:params}f).

\end{appendix}


\begin{thebibliography}{}
\expandafter\ifx\csname natexlab\endcsname\relax\def\natexlab#1{#1}\fi
\providecommand{\url}[1]{\href{#1}{#1}}
\providecommand{\dodoi}[1]{doi:~\href{http://doi.org/#1}{\nolinkurl{#1}}}
\providecommand{\doeprint}[1]{\href{http://ascl.net/#1}{\nolinkurl{http://ascl.net/#1}}}
\providecommand{\doarXiv}[1]{\href{https://arxiv.org/abs/#1}{\nolinkurl{https://arxiv.org/abs/#1}}}

\bibitem[{{Abramowicz} {et~al.}(1988){Abramowicz}, {Czerny}, {Lasota}, \&
  {Szuszkiewicz}}]{Abramowicz1988}
{Abramowicz}, M.~A., {Czerny}, B., {Lasota}, J.~P., \& {Szuszkiewicz}, E. 1988,
  \apj, 332, 646, \dodoi{10.1086/166683}

\bibitem[{{Chabrier}(2003)}]{Chabrier2003}
{Chabrier}, G. 2003, \pasp, 115, 763, \dodoi{10.1086/376392}

\bibitem[{{Davies} {et~al.}(2019){Davies}, {Hennawi}, \& {Eilers}}]{Davies2019}
{Davies}, F.~B., {Hennawi}, J.~F., \& {Eilers}, A.-C. 2019, \apjl, 884, L19,
  \dodoi{10.3847/2041-8213/ab42e3}

\bibitem[{{Decarli} {et~al.}(2022){Decarli}, {Pensabene}, {Venemans}, {Walter},
  {Banados}, {Bertoldi}, {Carilli}, {Cox}, {Fan}, {Farina}, {Ferkinhoff},
  {Groves}, {Li}, {Mazzucchelli}, {Neri}, {Riechers}, {Uzgil}, {Wang}, {Wang},
  {Weiss}, {Winters}, \& {Yang}}]{Decarli2022}
{Decarli}, R., {Pensabene}, A., {Venemans}, B., {et~al.} 2022, arXiv e-prints,
  arXiv:2203.03658.
\newblock \doarXiv{2203.03658}

\bibitem[{{Ding} {et~al.}(2020){Ding}, {Silverman}, {Treu}, {Schulze},
  {Schramm}, {Birrer}, {Park}, {Jahnke}, {Bennert}, {Kartaltepe}, {Koekemoer},
  {Malkan}, \& {Sanders}}]{Ding2020a}
{Ding}, X., {Silverman}, J., {Treu}, T., {et~al.} 2020, \apj, 888, 37,
  \dodoi{10.3847/1538-4357/ab5b90}

\bibitem[{{Eilers} {et~al.}(2021){Eilers}, {Hennawi}, {Davies}, \&
  {Simcoe}}]{Eilers2021}
{Eilers}, A.-C., {Hennawi}, J.~F., {Davies}, F.~B., \& {Simcoe}, R.~A. 2021,
  arXiv e-prints, arXiv:2106.04586.
\newblock \doarXiv{2106.04586}

\bibitem[{{Fan} {et~al.}(2000){Fan}, {White}, {Davis}, {Becker}, {Strauss},
  {Haiman}, {Schneider}, {Gregg}, {Gunn}, {Knapp}, {Lupton}, {Anderson},
  {Anderson}, {Annis}, {Bahcall}, {Boroski}, {Brunner}, {Chen}, {Connolly},
  {Csabai}, {Doi}, {Fukugita}, {Hennessy}, {Hindsley}, {Ichikawa},
  {Ivezi{\'c}}, {Loveday}, {Meiksin}, {McKay}, {Munn}, {Newberg}, {Nichol},
  {Okamura}, {Pier}, {Sekiguchi}, {Shimasaku}, {Stoughton}, {Szalay},
  {Szokoly}, {Thakar}, {Vogeley}, \& {York}}]{Fan2000}
{Fan}, X., {White}, R.~L., {Davis}, M., {et~al.} 2000, \aj, 120, 1167,
  \dodoi{10.1086/301534}

\bibitem[{{Genzel} {et~al.}(2017){Genzel}, {F{\"o}rster Schreiber},
  {{\"U}bler}, {Lang}, {Naab}, {Bender}, {Tacconi}, {Wisnioski}, {Wuyts},
  {Alexander}, {Beifiori}, {Belli}, {Brammer}, {Burkert}, {Carollo}, {Chan},
  {Davies}, {Fossati}, {Galametz}, {Genel}, {Gerhard}, {Lutz}, {Mendel},
  {Momcheva}, {Nelson}, {Renzini}, {Saglia}, {Sternberg}, {Tacchella},
  {Tadaki}, \& {Wilman}}]{Genzel2017}
{Genzel}, R., {F{\"o}rster Schreiber}, N.~M., {{\"U}bler}, H., {et~al.} 2017,
  \nat, 543, 397, \dodoi{10.1038/nature21685}

\bibitem[{{Grazian} {et~al.}(2015){Grazian}, {Fontana}, {Santini}, {Dunlop},
  {Ferguson}, {Castellano}, {Amorin}, {Ashby}, {Barro}, {Behroozi}, {Boutsia},
  {Caputi}, {Chary}, {Dekel}, {Dickinson}, {Faber}, {Fazio}, {Finkelstein},
  {Galametz}, {Giallongo}, {Giavalisco}, {Grogin}, {Guo}, {Kocevski},
  {Koekemoer}, {Koo}, {Lee}, {Lu}, {Merlin}, {Mobasher}, {Nonino}, {Papovich},
  {Paris}, {Pentericci}, {Reddy}, {Renzini}, {Salmon}, {Salvato}, {Sommariva},
  {Song}, \& {Vanzella}}]{Grazian2015}
{Grazian}, A., {Fontana}, A., {Santini}, P., {et~al.} 2015, \aap, 575, A96,
  \dodoi{10.1051/0004-6361/201424750}

\bibitem[{{Habouzit} {et~al.}(2022){Habouzit}, {Onoue}, {Ba{\~n}ados},
  {Neeleman}, {Angl{\'e}s-Alc{\'a}zar}, {Walter}, {Pillepich}, {Dav{\'e}},
  {Jahnke}, \& {Dubois}}]{Habouzit2022}
{Habouzit}, M., {Onoue}, M., {Ba{\~n}ados}, E., {et~al.} 2022, \mnras, 511,
  3751, \dodoi{10.1093/mnras/stac225}

\bibitem[{{H{\"a}ring} \& {Rix}(2004)}]{Haring2004}
{H{\"a}ring}, N., \& {Rix}, H.-W. 2004, \apjl, 604, L89, \dodoi{10.1086/383567}

\bibitem[{{Homayouni} {et~al.}(2020){Homayouni}, {Trump}, {Grier}, {Horne},
  {Shen}, {Brandt}, {Dawson}, {Alvarez}, {Green}, {Hall}, {Hern{\'a}ndez
  Santisteban}, {Ho}, {Kinemuchi}, {Kochanek}, {Li}, {Peterson}, {Schneider},
  {Starkey}, {Bizyaev}, {Pan}, {Oravetz}, \& {Simmons}}]{Homayouni2020}
{Homayouni}, Y., {Trump}, J.~R., {Grier}, C.~J., {et~al.} 2020, \apj, 901, 55,
  \dodoi{10.3847/1538-4357/ababa9}

\bibitem[{{Inayoshi} {et~al.}(2019){Inayoshi}, {Ichikawa}, {Ostriker}, \&
  {Kuiper}}]{Inayoshi2019disk}
{Inayoshi}, K., {Ichikawa}, K., {Ostriker}, J.~P., \& {Kuiper}, R. 2019,
  \mnras, 486, 5377, \dodoi{10.1093/mnras/stz1189}

\bibitem[{{Inayoshi} {et~al.}(2022){Inayoshi}, {Nakatani}, {Toyouchi},
  {Hosokawa}, {Kuiper}, \& {Onoue}}]{Inayoshi2022}
{Inayoshi}, K., {Nakatani}, R., {Toyouchi}, D., {et~al.} 2022, \apj, 927, 237,
  \dodoi{10.3847/1538-4357/ac4751}

\bibitem[{{Inayoshi} {et~al.}(2020){Inayoshi}, {Visbal}, \&
  {Haiman}}]{Inayoshi2020}
{Inayoshi}, K., {Visbal}, E., \& {Haiman}, Z. 2020, \araa, 58, 27,
  \dodoi{10.1146/annurev-astro-120419-014455}

\bibitem[{{Izumi} {et~al.}(2019){Izumi}, {Onoue}, {Matsuoka}, {Nagao},
  {Strauss}, {Imanishi}, {Kashikawa}, {Fujimoto}, {Kohno}, {Toba}, {Umehata},
  {Goto}, {Ueda}, {Shirakata}, {Silverman}, {Greene}, {Harikane}, {Hashimoto},
  {Ikarashi}, {Iono}, {Iwasawa}, {Lee}, {Minezaki}, {Nakanishi}, {Tamura},
  {Tang}, \& {Taniguchi}}]{Izumi2019}
{Izumi}, T., {Onoue}, M., {Matsuoka}, Y., {et~al.} 2019, \pasj, 71, 111,
  \dodoi{10.1093/pasj/psz096}

\bibitem[{{Izumi} {et~al.}(2021){Izumi}, {Matsuoka}, {Fujimoto}, {Onoue},
  {Strauss}, {Umehata}, {Imanishi}, {Kohno}, {Kawaguchi}, {Kawamuro}, {Baba},
  {Nagao}, {Toba}, {Inayoshi}, {Silverman}, {Inoue}, {Ikarashi}, {Iwasawa},
  {Kashikawa}, {Hashimoto}, {Nakanishi}, {Ueda}, {Schramm}, {Lee}, \&
  {Suh}}]{Izumi2021}
{Izumi}, T., {Matsuoka}, Y., {Fujimoto}, S., {et~al.} 2021, \apj, 914, 36,
  \dodoi{10.3847/1538-4357/abf6dc}

\bibitem[{{Jahnke} {et~al.}(2009){Jahnke}, {Bongiorno}, {Brusa}, {Capak},
  {Cappelluti}, {Cisternas}, {Civano}, {Colbert}, {Comastri}, {Elvis},
  {Hasinger}, {Ilbert}, {Impey}, {Inskip}, {Koekemoer}, {Lilly}, {Maier},
  {Merloni}, {Riechers}, {Salvato}, {Schinnerer}, {Scoville}, {Silverman},
  {Taniguchi}, {Trump}, \& {Yan}}]{Jahnke2009}
{Jahnke}, K., {Bongiorno}, A., {Brusa}, M., {et~al.} 2009, \apjl, 706, L215,
  \dodoi{10.1088/0004-637X/706/2/L215}

\bibitem[{{Kelly} \& {Shen}(2013)}]{Kelly2013}
{Kelly}, B.~C., \& {Shen}, Y. 2013, \apj, 764, 45,
  \dodoi{10.1088/0004-637X/764/1/45}

\bibitem[{{King} \& {Pounds}(2015)}]{King2015}
{King}, A., \& {Pounds}, K. 2015, \araa, 53, 115,
  \dodoi{10.1146/annurev-astro-082214-122316}

\bibitem[{{Kormendy} \& {Ho}(2013)}]{Kormendy2013}
{Kormendy}, J., \& {Ho}, L.~C. 2013, \araa, 51, 511,
  \dodoi{10.1146/annurev-astro-082708-101811}

\bibitem[{{Lauer} {et~al.}(2007){Lauer}, {Tremaine}, {Richstone}, \&
  {Faber}}]{Lauer2007}
{Lauer}, T.~R., {Tremaine}, S., {Richstone}, D., \& {Faber}, S.~M. 2007, \apj,
  670, 249, \dodoi{10.1086/522083}

\bibitem[{{Li} {et~al.}(2021){Li}, {Silverman}, {Ding}, {Strauss}, {Goulding},
  {Schramm}, {Yesuf}, {Sun}, {Xue}, {Birrer}, {Shi}, {Toba}, {Nagao}, \&
  {Imanishi}}]{Li2021}
{Li}, J., {Silverman}, J.~D., {Ding}, X., {et~al.} 2021, \apj, 922, 142,
  \dodoi{10.3847/1538-4357/ac2301}

\bibitem[{{Marshall} {et~al.}(2021){Marshall}, {Wyithe}, {Windhorst}, {Matteo},
  {Ni}, {Wilkins}, {Croft}, \& {Mechtley}}]{Marshall2021}
{Marshall}, M.~A., {Wyithe}, J. S.~B., {Windhorst}, R.~A., {et~al.} 2021,
  \mnras, 506, 1209, \dodoi{10.1093/mnras/stab1763}

\bibitem[{{Matsuoka} {et~al.}(2016){Matsuoka}, {Onoue}, {Kashikawa}, {Iwasawa},
  {Strauss}, {Nagao}, {Imanishi}, {Niida}, {Toba}, {Akiyama}, {Asami}, {Bosch},
  {Foucaud}, {Furusawa}, {Goto}, {Gunn}, {Harikane}, {Ikeda}, {Kawaguchi},
  {Kikuta}, {Komiyama}, {Lupton}, {Minezaki}, {Miyazaki}, {Morokuma},
  {Murayama}, {Nishizawa}, {Ono}, {Ouchi}, {Price}, {Sameshima}, {Silverman},
  {Sugiyama}, {Tait}, {Takada}, {Takata}, {Tanaka}, {Tang}, \&
  {Utsumi}}]{Matsuoka2016}
{Matsuoka}, Y., {Onoue}, M., {Kashikawa}, N., {et~al.} 2016, \apj, 828, 26,
  \dodoi{10.3847/0004-637X/828/1/26}

\bibitem[{{McKinney} {et~al.}(2021){McKinney}, {Hayward}, {Rosenthal},
  {Mart{\'\i}nez-Galarza}, {Pope}, {Sajina}, \& {Smith}}]{McKinney2021}
{McKinney}, J., {Hayward}, C.~C., {Rosenthal}, L.~J., {et~al.} 2021, \apj, 921,
  55, \dodoi{10.3847/1538-4357/ac185f}

\bibitem[{{Merloni} {et~al.}(2010){Merloni}, {Bongiorno}, {Bolzonella},
  {Brusa}, {Civano}, {Comastri}, {Elvis}, {Fiore}, {Gilli}, {Hao}, {Jahnke},
  {Koekemoer}, {Lusso}, {Mainieri}, {Mignoli}, {Miyaji}, {Renzini}, {Salvato},
  {Silverman}, {Trump}, {Vignali}, {Zamorani}, {Capak}, {Lilly}, {Sanders},
  {Taniguchi}, {Bardelli}, {Carollo}, {Caputi}, {Contini}, {Coppa}, {Cucciati},
  {de la Torre}, {de Ravel}, {Franzetti}, {Garilli}, {Hasinger}, {Impey},
  {Iovino}, {Iwasawa}, {Kampczyk}, {Kneib}, {Knobel}, {Kova{\v{c}}},
  {Lamareille}, {Le Borgne}, {Le Brun}, {Le F{\`e}vre}, {Maier}, {Pello},
  {Peng}, {Perez Montero}, {Ricciardelli}, {Scodeggio}, {Tanaka}, {Tasca},
  {Tresse}, {Vergani}, \& {Zucca}}]{Merloni2010}
{Merloni}, A., {Bongiorno}, A., {Bolzonella}, M., {et~al.} 2010, \apj, 708,
  137, \dodoi{10.1088/0004-637X/708/1/137}

\bibitem[{{Molina} {et~al.}(2021){Molina}, {J.}, {Wang}, {R.}, {Shangguan},
  {J.}, {Ho}, {C.}, {Bauer}, {E.}, {Treister}, {E.}, {Shao}, \&
  {Y}}]{Molina2021}
{Molina}, {J.}, {Wang}, {et~al.} 2021, arXiv e-prints, arXiv:2101.00764.
\newblock \doarXiv{2101.00764}

\bibitem[{{Murphy} {et~al.}(2011){Murphy}, {Condon}, {Schinnerer}, {Kennicutt},
  {Calzetti}, {Armus}, {Helou}, {Turner}, {Aniano}, {Beir{\~a}o}, {Bolatto},
  {Brandl}, {Croxall}, {Dale}, {Donovan Meyer}, {Draine}, {Engelbracht},
  {Hunt}, {Hao}, {Koda}, {Roussel}, {Skibba}, \& {Smith}}]{Murphy2011}
{Murphy}, E.~J., {Condon}, J.~J., {Schinnerer}, E., {et~al.} 2011, \apj, 737,
  67, \dodoi{10.1088/0004-637X/737/2/67}

\bibitem[{{Neeleman} {et~al.}(2021){Neeleman}, {Novak}, {Venemans}, {Walter},
  {Decarli}, {Kaasinen}, {Schindler}, {Ba{\~n}ados}, {Carilli}, {Drake}, {Fan},
  \& {Rix}}]{Neeleman2021}
{Neeleman}, M., {Novak}, M., {Venemans}, B.~P., {et~al.} 2021, \apj, 911, 141,
  \dodoi{10.3847/1538-4357/abe70f}

\bibitem[{{Onoue} {et~al.}(2019){Onoue}, {Kashikawa}, {Matsuoka}, {Kato},
  {Izumi}, {Nagao}, {Strauss}, {Harikane}, {Imanishi}, {Ito}, {Iwasawa},
  {Kawaguchi}, {Lee}, {Noboriguchi}, {Suh}, {Tanaka}, \& {Toba}}]{Onoue2019}
{Onoue}, M., {Kashikawa}, N., {Matsuoka}, Y., {et~al.} 2019, \apj, 880, 77,
  \dodoi{10.3847/1538-4357/ab29e9}

\bibitem[{{Onoue} {et~al.}(2021){Onoue}, {Matsuoka}, {Kashikawa}, {Strauss},
  {Iwasawa}, {Izumi}, {Nagao}, {Asami}, {Fujimoto}, {Harikane}, {Hashimoto},
  {Imanishi}, {Lee}, {Shibuya}, \& {Toba}}]{Onoue2021}
{Onoue}, M., {Matsuoka}, Y., {Kashikawa}, N., {et~al.} 2021, \apj, 919, 61,
  \dodoi{10.3847/1538-4357/ac0f07}

\bibitem[{{Pensabene} {et~al.}(2020){Pensabene}, {Carniani}, {Perna}, {Cresci},
  {Decarli}, {Maiolino}, \& {Marconi}}]{Pensabene2020}
{Pensabene}, A., {Carniani}, S., {Perna}, M., {et~al.} 2020, \aap, 637, A84,
  \dodoi{10.1051/0004-6361/201936634}

\bibitem[{{Richards} {et~al.}(2006){Richards}, {Lacy}, {Storrie-Lombardi},
  {Hall}, {Gallagher}, {Hines}, {Fan}, {Papovich}, {Vanden Berk}, {Trammell},
  {Schneider}, {Vestergaard}, {York}, {Jester}, {Anderson}, {Budav{\'a}ri}, \&
  {Szalay}}]{Richards2006}
{Richards}, G.~T., {Lacy}, M., {Storrie-Lombardi}, L.~J., {et~al.} 2006, \apjs,
  166, 470, \dodoi{10.1086/506525}

\bibitem[{{Schramm} \& {Silverman}(2013)}]{Schramm2013}
{Schramm}, M., \& {Silverman}, J.~D. 2013, \apj, 767, 13,
  \dodoi{10.1088/0004-637X/767/1/13}

\bibitem[{{Schulze} \& {Wisotzki}(2014)}]{Schulze2014}
{Schulze}, A., \& {Wisotzki}, L. 2014, \mnras, 438, 3422,
  \dodoi{10.1093/mnras/stt2457}

\bibitem[{{Shakura} \& {Sunyaev}(1973)}]{Shakura1973}
{Shakura}, N.~I., \& {Sunyaev}, R.~A. 1973, \aap, 24, 337

\bibitem[{{Shen}(2013)}]{Shen2013}
{Shen}, Y. 2013, Bulletin of the Astronomical Society of India, 41, 61.
\newblock \doarXiv{1302.2643}

\bibitem[{{Shen} {et~al.}(2019){Shen}, {Wu}, {Jiang}, {Ba{\~n}ados}, {Fan},
  {Ho}, {Riechers}, {Strauss}, {Venemans}, {Vestergaard}, {Walter}, {Wang},
  {Willott}, {Wu}, \& {Yang}}]{Shen2019}
{Shen}, Y., {Wu}, J., {Jiang}, L., {et~al.} 2019, \apj, 873, 35,
  \dodoi{10.3847/1538-4357/ab03d9}

\bibitem[{{Speagle} {et~al.}(2014){Speagle}, {Steinhardt}, {Capak}, \&
  {Silverman}}]{Speagle2014}
{Speagle}, J.~S., {Steinhardt}, C.~L., {Capak}, P.~L., \& {Silverman}, J.~D.
  2014, \apjs, 214, 15, \dodoi{10.1088/0067-0049/214/2/15}

\bibitem[{{Sun} {et~al.}(2015){Sun}, {Trump}, {Brandt}, {Luo}, {Alexander},
  {Jahnke}, {Rosario}, {Wang}, \& {Xue}}]{Sun2015}
{Sun}, M., {Trump}, J.~R., {Brandt}, W.~N., {et~al.} 2015, \apj, 802, 14,
  \dodoi{10.1088/0004-637X/802/1/14}

\bibitem[{{Tacconi} {et~al.}(2018){Tacconi}, {Genzel}, {Saintonge}, {Combes},
  {Garc{\'\i}a-Burillo}, {Neri}, {Bolatto}, {Contini}, {F{\"o}rster Schreiber},
  {Lilly}, {Lutz}, {Wuyts}, {Accurso}, {Boissier}, {Boone}, {Bouch{\'e}},
  {Bournaud}, {Burkert}, {Carollo}, {Cooper}, {Cox}, {Feruglio}, {Freundlich},
  {Herrera-Camus}, {Juneau}, {Lippa}, {Naab}, {Renzini}, {Salome}, {Sternberg},
  {Tadaki}, {{\"U}bler}, {Walter}, {Weiner}, \& {Weiss}}]{Tacconi2018}
{Tacconi}, L.~J., {Genzel}, R., {Saintonge}, A., {et~al.} 2018, \apj, 853, 179,
  \dodoi{10.3847/1538-4357/aaa4b4}

\bibitem[{{Trebitsch} {et~al.}(2019){Trebitsch}, {Volonteri}, \&
  {Dubois}}]{Trebitsch2019}
{Trebitsch}, M., {Volonteri}, M., \& {Dubois}, Y. 2019, \mnras, 487, 819,
  \dodoi{10.1093/mnras/stz1280}

\bibitem[{{Valentini} {et~al.}(2021){Valentini}, {Gallerani}, \&
  {Ferrara}}]{Valentini2021}
{Valentini}, M., {Gallerani}, S., \& {Ferrara}, A. 2021, \mnras, 507, 1,
  \dodoi{10.1093/mnras/stab1992}

\bibitem[{{Venemans} {et~al.}(2016){Venemans}, {Walter}, {Zschaechner},
  {Decarli}, {De Rosa}, {Findlay}, {McMahon}, \& {Sutherland}}]{Venemans2016}
{Venemans}, B.~P., {Walter}, F., {Zschaechner}, L., {et~al.} 2016, \apj, 816,
  37, \dodoi{10.3847/0004-637X/816/1/37}

\bibitem[{{Venemans} {et~al.}(2017){Venemans}, {Walter}, {Decarli},
  {Ba{\~n}ados}, {Hodge}, {Hewett}, {McMahon}, {Mortlock}, \&
  {Simpson}}]{Venemans2017}
{Venemans}, B.~P., {Walter}, F., {Decarli}, R., {et~al.} 2017, \apj, 837, 146,
  \dodoi{10.3847/1538-4357/aa62ac}

\bibitem[{{Vestergaard} \& {Osmer}(2009)}]{Vestergaard2009}
{Vestergaard}, M., \& {Osmer}, P.~S. 2009, \apj, 699, 800,
  \dodoi{10.1088/0004-637X/699/1/800}

\bibitem[{{Vito} {et~al.}(2018){Vito}, {Brandt}, {Yang}, {Gilli}, {Luo},
  {Vignali}, {Xue}, {Comastri}, {Koekemoer}, {Lehmer}, {Liu}, {Paolillo},
  {Ranalli}, {Schneider}, {Shemmer}, {Volonteri}, \& {Wang}}]{Vito2018}
{Vito}, F., {Brandt}, W.~N., {Yang}, G., {et~al.} 2018, \mnras, 473, 2378,
  \dodoi{10.1093/mnras/stx2486}

\bibitem[{{Volonteri} \& {Reines}(2016)}]{Volonteri2016}
{Volonteri}, M., \& {Reines}, A.~E. 2016, \apjl, 820, L6,
  \dodoi{10.3847/2041-8205/820/1/L6}

\bibitem[{{Volonteri} \& {Stark}(2011)}]{Volonteri2011}
{Volonteri}, M., \& {Stark}, D.~P. 2011, \mnras, 417, 2085,
  \dodoi{10.1111/j.1365-2966.2011.19391.x}

\bibitem[{{Wang} {et~al.}(2013){Wang}, {Wagg}, {Carilli}, {Walter}, {Lentati},
  {Fan}, {Riechers}, {Bertoldi}, {Narayanan}, {Strauss}, {Cox}, {Omont},
  {Menten}, {Knudsen}, {Neri}, \& {Jiang}}]{Wang2013}
{Wang}, R., {Wagg}, J., {Carilli}, C.~L., {et~al.} 2013, \apj, 773, 44,
  \dodoi{10.1088/0004-637X/773/1/44}

\bibitem[{{Yang} {et~al.}(2020){Yang}, {Shen}, {Chen}, {Liu}, {Annis}, {Avila},
  {Bertin}, {Brooks}, {Buckley-Geer}, {Carnero Rosell}, {Carrasco Kind},
  {Carretero}, {da Costa}, {Desai}, {Thomas Diehl}, {Doel}, {Frieman},
  {Garcia-Bellido}, {Gaztanaga}, {Gerdes}, {Gruen}, {Gruendl}, {Gschwend},
  {Gutierrez}, {Hollowood}, {Honscheid}, {Hoyle}, {James}, {Krause}, {Kuehn},
  {Lidman}, {Lima}, {Maia}, {Marshall}, {Martini}, {Menanteau}, {Miquel},
  {Plazas Malag{\'o}n}, {Sanchez}, {Scarpine}, {Schindler}, {Schubnell},
  {Serrano}, {Sevilla}, {Smith}, {Soares-Santos}, {Sobreira}, {Suchyta},
  {Swanson}, {Tarle}, {Vikram}, \& {Walker}}]{Yang2020}
{Yang}, Q., {Shen}, Y., {Chen}, Y.-C., {et~al.} 2020, \mnras, 493, 5773,
  \dodoi{10.1093/mnras/staa645}

\end{thebibliography}
\bibliographystyle{aasjournal}

\end{document}